\documentclass[a4paper,11pt]{article}
\pdfoutput=1 
\usepackage{jinstpub} 
                     
\usepackage{lineno}

\begin{document}
\title{Study of muon pair production from positron annihilation at threshold
energy}


\author[a,b]{N. Amapane,}
\author[c]{M. Antonelli,}
\author[d]{F. Anulli,}
\author[e,f]{G. Ballerini,}
\author[g]{L. Bandiera,}
\author[b]{N. Bartosik,}
\author[d]{M. Bauce,}
\author[h]{A. Bertolin,}
\author[b]{C. Biino,}
\author[c]{O. R. Blanco-Garcia,}
\author[c]{M. Boscolo,}
\author[e,f]{C. Brizzolari,}
\author[a,b]{A. Cappati,}
\author[i]{M. Casarsa,}
\author[l,d]{G. Cavoto,}
\author[d]{F. Collamati,}
\author[a,b]{G. Cotto,}
\author[h]{C. Curatolo,}
\author[m]{R. Di Nardo,}
\author[h]{F. Gonella,}
\author[n,h]{S. Hoh,}
\author[c]{M. Iafrati,}
\author[d]{F. Iacoangeli,}
\author[b]{B. Kiani,}
\author[n,h]{D. Lucchesi,}
\author[e,f]{V. Mascagna,}
\author[n,h]{A. Paccagnella,}
\author[b]{N. Pastrone,}
\author[n,h]{J. Pazzini,}
\author[b]{M. Pelliccioni,}
\author[c]{B. Ponzio,}
\author[e,f]{M. Prest,}
\author[c]{M. Ricci,}
\author[n,h]{R. Rossin,}
\author[c]{M. Rotondo,}
\author[a,b]{O. Sans Planell,}
\author[h]{L. Sestini,}
\author[e,f]{M. Soldani,}
\author[o]{A. Triossi,}
\author[f]{E. Vallazza,}
\author[h]{S. Ventura,}
\author[n,h]{M. Zanetti.}

\affiliation[a]{Universit\`a degli Studi di Torino - Torino, Italy}
\affiliation[b]{INFN Sezione di Torino - Torino, Italy}
\affiliation[c]{INFN Laboratori Nazionali di Frascati - Frascati, Italy}
\affiliation[d]{INFN Sezione di Roma - Rome, Italy}
\affiliation[e]{Universit\`a degli Studi dell'Insubria - Como, Italy}
\affiliation[f]{INFN Sezione di Milano Bicocca - Milan, Italy}
\affiliation[g]{INFN Sezione di Ferrara - Ferrara, Italy}
\affiliation[h]{INFN Sezione di Padova - Padova, Italy}
\affiliation[i]{INFN Sezione di Trieste - Trieste, Italy}
\affiliation[l]{Universit\`a di Roma La Sapienza - Rome, Italy}
\affiliation[m]{CERN - Geneva, Switzerland}
\affiliation[n]{Universit\`a di Padova - Padova, Italy}
\affiliation[o]{Institut Pluridisciplinaire Hubert Curien, Strasbourg, France}

\emailAdd{bertolin@pd.infn.it}

\abstract{
\noindent
The muon collider represents one of the most promising solutions for a future machine exploring the high energy frontier, but several challenges due to the 2.2 $\mu$sec muon lifetime at rest have
to be carefully considered. The LEMMA project is investigating  the possibility of producing low emittance muon/antimuon pairs from the e$^+$e$^-$ annihilation process at threshold energy,
resulting in small transverse emittance beams without any additional beam cooling.
However most of the measurements available are performed at higher $\sqrt{s}$ values.
It is therefore necessary to measure muons production in positron annihilation at threshold energy
and compare the experimental results with the predictions in this specific energy regime.
Apart from being a topic of physical interest by itself, these near to threshold measurements 
can have a sizeable impact on the estimation of the ultimate luminosity achievable in a muon collider with the LEMMA injection scheme.
}


\maketitle

\section{Introduction}

The muon collider represents one of the best solutions for a future machine
at the energy frontier because it can provide, still using elementary particles, 
a center of mass energy much higher that any electron collider.
A muon collider facility has been studied and designed by the MAP 
project \cite{MAP}. This study demonstrated that a muon collider is feasible 
up to 6 TeV hence exploring the multi--TeV energy frontier with the possibility 
to study Higgs boson properties.
However a muon collider has to face several challenges.
The muon lifetime of 2.2 $\mu$sec at rest requires a fast accelerator chain.
The production of low emittance muon/antimuon beams to be fed into the accelerator 
complex imposes the use of fast muon cooling techniques \cite{MICE} when muons are produced
by decay of pions.

The LEMMA project~\cite{NIM} aims to study the possibility of producing muons
from the e$^+$e$^-$ annihilation process. The idea is to use a high intensity positron beam, above the production energy threshold at 43.7~GeV, that impinges on a fixed target. In this way muons are produced with a small divergence, resulting in a small transverse emittance that could avoid the need of beam cooling.

While the leading-order QED cross section $e^+ e^- \rightarrow \mu^+ \mu^-$ is well established, near threshold the lowest order radiative corrections and effects due to the Coulomb interaction in the final state become essential and cannot be neglected \cite{coulomb}. Experimental data in this specific regime are not frequent as most of the measurements are performed at higher $\sqrt{s}$ values \cite{thomson}. It is therefore necessary to measure the cross section and the
$\mu^+ \mu^-$ kinematical properties for several values of the center of mass energy near threshold to probe such predictions. 
Apart from being a topic of physical interest by itself, with impact on g-2 measurements \cite{coulomb}, these near to threshold corrections can have a sizeable impact on the final luminosity achievable in a muon collider with the LEMMA injection scheme. 
Moreover the $e^+ e^- \rightarrow \mu^+ \mu^-$ cross section as implemented in the Geant4 \cite{geant4} simulation is being used to tune the LEMMA injection scheme parameters, 
but near-threshold corrections are not applied in Geant4. Moreover, the Geant4 implementation has not been experimentally tested in this particular regime.

The aim of these measurements is therefore the study of muon pair production in the described conditions and the comparison of the measured muon emittance with simulation results.
Section 2 presents the experimental setup implemented in a test beam campaign performed at CERN in Summer 2018. Section 3 reports the analytical computations as well as a detailed Monte Carlo simulation that has been developed to get predictions that could be compared to data. The data taking and analysis strategy is described in Section 4. Section 5 gives the
measurements of several physical quantities related to the $\mu^+ \mu^-$ system and compares
them to simulations.
Conclusions are given in Section 6.

\section{Experimental setup} 

\begin{figure}[ht]
  \begin{center}
    \includegraphics[width=\linewidth]{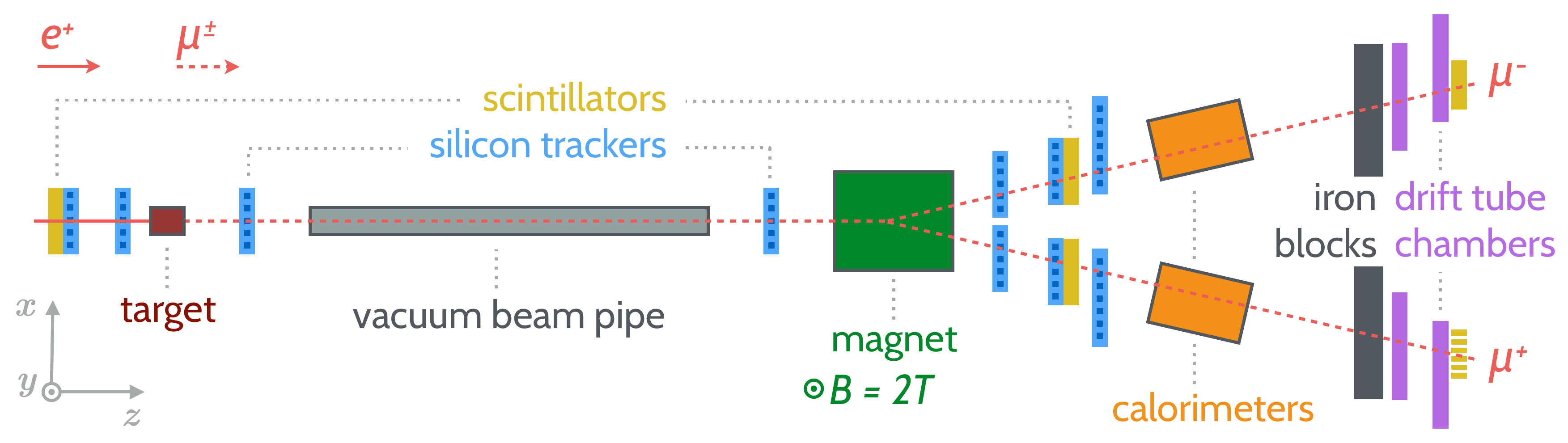}
  \end{center}
  \caption{Schematic view of the test beam experimental setup as used in the testbeam of August-September 2018. The $x$-$z$ plane view is shown, with the $x$ and $z$ axes as defined in the text. All the components of the setup are marked with a corresponding colour in the figure.
  The scintillator on the $\mu^+$ arm downstream of the drift tube chambers was present only in the September 2018 setup and hence represented graphically in a different way.}
  \label{fig:tb_setup}
\end{figure}

The experimental setup was designed to measure with high precision trajectories and momenta of the two final state muons as well as the direction of interacting positrons.
The layout of the setup is schematically shown in Fig.~\ref{fig:tb_setup}, with the right-handed coordinate system defined by the 
$z$-axis pointing along the direction of the positron beam and the $y$-axis pointing to the roof of the experimental hall.
The total length of the apparatus in the $z$ direction was about 23 m.

Initially the positron beam passed through a fast-response plastic scintillator and a pair of silicon microstrip sensors before hitting the target. The silicon sensors allowed a measurement of the direction and position of the positron beam on the target, while the scintillator was used for trigger purposes.
In this experiment all the silicon sensors were formed by two layers of microstrips with orthogonal directions, and allowed a measurement of the hit positions and the corresponding pulse height in the $x$ $y$ plane.
The silicon sensors upstream of the target were 2 $\times$ 2 cm$^2$ in size and provided position measurements with a microstrip pitch of 50 $\mu$m~\cite{T1T2T3}.
In the following the $z$-position of the first silicon detector was considered as the origin of the $z$-axis ($z=0$). 
The measured $z$ positions of the detectors and the target of the August 2018 setup are given in the following. They are compatible within a few centimeters with the September 2018 setup. 
The second silicon sensor was placed at $z=359$ cm, while the center of the target was located at $z=458$ cm.

Muon pairs produced in the target material passed through a vacuum beam pipe and another pair of silicon sensors, one before ($z=467$ cm) and one after the pipe ($z=1410$ cm), which measured the direction of the muons before passing through the 2~T magnetic field created by a dipole magnet.
The two silicon microstrip sensors downstream of the target and upstream of the magnet were 9.5 $\times$ 9.5 cm$^2$ in size and had a pitch of 242 $\mu$m~\cite{C1-C10}.
The center of the dipole magnet was at $z=1591$ cm and the field extended in a region of approximately $\pm 1000$ cm.

Downstream of the magnet the paths of the muons diverged in two arms.
Muon momenta were calculated using position measurements upstream and downstream of the bending magnet.
In each arm a muon passed through three layers of silicon microstrip sensors ($z=$1791 cm, $z=$1913 cm and $z=$2042 cm) followed by a calorimeter and by two layers of drift tube (DT) muon chambers (first layer at $z=$2311 cm).
The two silicon sensors upstream and closer to the magnet were 8 $\times$ 8 cm$^2$ in size and had a pitch of 228 $\mu$m, while the silicon sensor in front of the DT chamber was 18 $\times$ 18 cm$^2$ with a pitch of 456 $\mu$m ~\cite{C1-C10}.

The DT chambers employed the same technology used by the CMS experiment at LHC \cite{ref:CMS-DT}.
The DT chambers consisted of four layers of wires each, providing up to 8 hits in the $x$ $z$ plane in each arm.
The expected spatial resolution along the $x$-axis was about 150 $\mu$m.

Positrons were expected to deposit most of their energy in the calorimeter.
Additional leakage was absorbed by the iron shielding placed downstream so that only $\mu^\pm$ tracks were expected to reach the DT chambers.
Each calorimeter consisted of a lead-glass (PbWO$_{4}$) section followed by a Cherenkov section, to differentiate between electron and muon tracks.

Finally, a plastic scintillator was positioned after the last DT chamber, 
only in the $\mu^-$ arm in the August run, so that a total of four scintillators were present. An additional scintillator was included in the September run in the $\mu^+$ arm as well, for a total of five scintillators.
A coincidence between these plastic scintillators served as trigger for the shared silicon sensors and calorimeters data acquisition system (DAQ).
The DT chambers were using an independent trigger-less DAQ system with an acquisition rate of 40 MHz.
The trigger signal from the scintillators was shared between the two DAQ systems for offline synchronisation and event building.

\section{Simulations}

\subsection{Analytical computations}
\label{sec:analytical}

Analytical computations have been performed to get predictions for the emittance value of the muons produced by the process $e^+e^- \rightarrow \mu^+ \mu^-$ in a plane perpendicular to the $z$-axis at the $z$-coordinate corresponding to the target end position, taken as the reference plane.
The ad--hoc developed {\tt EEMUMU} code generates the muons by taking into account the incoming positron beam features (spatial and angular distributions of the positrons at the entrance of the target) and the target material.
The lowest order radiative corrections and effects due to the Coulomb interaction in the final state described in the introduction have not been taken into account.
The code has been benchmarked against {\tt Babayaga} \cite{CarloniCalame:2003yt} and {\tt Whizard} \cite{Kilian:2007gr} in case of an ideal (with no angular and energy spread)
positron beam impinging on an empty target. To match as accurately as possible the experimental conditions, a positron beam characterized by a flat distribution in a square of $2 \times 2$ cm$^2$ matching the shape of the silicon detectors upstream of the target, a gaussian distributed angular spread of $335$ $\mu$rad with a cutoff at 
$\pm$ 450 $\mu$rad and an energy of $45$ GeV, impinging on a cylindrical beryllium target $6$ cm long has been simulated. 
This resulted in an expected geometrical emittance of $3221$ nm $\times$ rad,
computed according to the trace--space prescription of \cite{emittance}:
\begin{equation}
\epsilon=\sqrt{(\langle x(\mu)^2 \rangle - \langle x(\mu) \rangle ^2)
(\langle x'(\mu)^2 \rangle - \langle x(\mu)' \rangle ^2) - 
(\langle x(\mu) \rangle \langle x(\mu)' \rangle - \langle x(\mu)x(\mu)' \rangle )^2}
\label{eq:emittance}
\end{equation}
where the variables $x(\mu)$ and $x(\mu)'$ are the position and angle of
the muons at the reference plane and
$\langle x(\mu) \rangle$, $\langle x(\mu)^2 \rangle$, 
$\langle x(\mu)' \rangle$, $\langle x(\mu)'^2 \rangle$ and $\langle x(\mu)x(\mu)' \rangle$ are the mean values of the corresponding quantities.
This analytical computation does not take into account detailed muon propagation effects in the material, apart from a parametrization of multiple scattering, and track reconstruction effects but this result is a useful cross check for the corresponding values found from both Monte Carlo simulation and data. 

\subsection{Monte Carlo simulations}

The whole experimental setup has been implemented in the Geant4 simulation toolkit \cite{geant4}.
All relevant volumes, silicon detectors, calorimeters, muon chambers and iron shielding, have been simulated with their exact shape and material composition in order to correctly model 
energy loss and multiple scattering.
As far as primary particles are concerned, several options have been implemented: besides the possibility to simulate a 45 GeV positron beam along the $z$-axis, with characteristics similar to the experimantal one, also the possibility to use a $\mu^+ \mu^-$ pairs input file in the HEPMC format has been implemented. This allowed the simulation of a sufficient amount of $e^+ e^- \rightarrow \mu^+ \mu^-$ events within a reasonable time, given the very low cross section of the annihilation process.
Similarly, the possibility to simulate Bhabha events via external input files has been implemented.
The magnetic field could be described in two ways: either with the measured field map provided for the specific magnet and running current used in the experiment or with an effective constant dipole value inside the magnet footprint. It was checked using simulations
that the two descriptions lead to very similar results and hence the second option was
used.
The output of the simulation is given on a particle basis recording each interaction in any of the volumes.

\section{Data taking and analysis strategy}

Two test beam data-taking campaigns have been conducted in August 2018 and September 2018, for approximately one week each time.
At the beginning of each period, calibration runs have been recorded in different configurations. In particular $\mu^+$ beams at energies of 22 GeV  without target and with both magnetic field directions have been used for alignment of silicon detectors and DT muon chambers in both arms.
In August 2018 physics runs were recorded with a 45 GeV positron beam impinging on a beryllium cylindrical target.
The beryllium target was 60 mm long and with a 40 mm diameter.
In September 2018 physics runs were recorded with positron beams of several energies (45, 46.5 and 49 GeV) impinging on the same beryllium target, and runs with a 45 GeV positron beam on carbon cylindrical targets with different length (60 mm and 20 mm) and same diameter (40 mm).
The positron beam had a pulsed shape with 4 spills per minute, each spill lasting 4.8 sec
with a typical intensity of 5 $\cdot$ 10$^6$ positrons. 
As anticipated in Sec. \ref{sec:analytical}, the spot size was $\sim$ 2 x 2 cm$^2$ with a
typical angular spread of $\sim$ 300 $\mu$rad. With the chosen collimators setting the
momentum spread was below 1.5 \% \cite{mom_spread}. The purity of the beam was in the range 95--99 \% \cite{Charitonidis}.


The goal of the data analysis is to identify $e^+e^- \rightarrow \mu^+ \mu^-$ events, measure the muon trajectories and the interacting positron direction in the $x$ $z$ plane and the muon momenta in the bending plane.
The main background process was the Bhabbha scattering $e^+e^- \rightarrow e^+ e^-$. Most of the electrons did not have enough energy to pass through the calorimeters, therefore events with two muons, one for each arm, were identified by requiring one track per DT muon chamber. DT tracks were used as seeds for the pattern recognition. Hits were added starting from the DT chambers and moving backwards to the other detectors with the following procedure:
\begin{enumerate}
    \item for each arm, two or three hits in the silicon detectors downstream to the magnet were selected if they formed, in the $x$ $z$ plane, a straight line with the corresponding DT track; fits to these hits were performed to obtain one downstream muon track per arm in the $x$ $z$ plane;
    \item a preliminary estimation of $\mu^+$ and $\mu^-$ momenta was obtained using the angles formed by downstream tracks and $z$-axis;
    \item $\mu^+$ and $\mu^-$ $x$-positions in the silicon detector before the magnet were extrapolated propagating the downstream tracks through the magnetic field, according to their estimated momenta; the two hits nearest to the extrapolated $\mu^+$ and $\mu^-$ positions were added to the tracks;
    \item a global fit that involved all the already selected hits was performed to add the best $\mu^+$ and $\mu^-$ hits in the first silicon detector after the target. Spline functions (straight line-parabola-straight line) were used to describe the muon trajectories in the $x$ $z$ plane at this stage;
    \item selected $\mu^+$ and $\mu^-$ tracks were then re-fitted to obtain the $x$ $z$ trajectories and the measured momenta. A cut on the goodness of fit was applied to remove the combinatorial background.
\end{enumerate}
The muon reconstruction algorithm was validated in the calibration runs, where muons were reconstructed and their measured momenta were compared with the nominal known value. Calibration runs were also used to verify that the electron mis-identification in the muon chambers was negligible. 

\begin{figure}[htb]
  \begin{center}
    \includegraphics[width=0.45\linewidth]{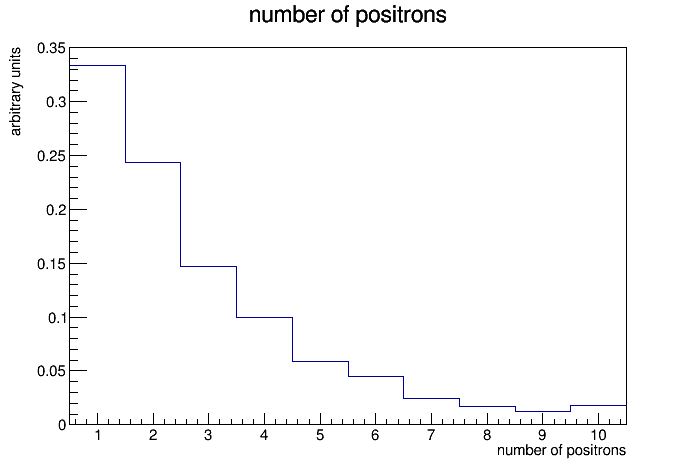}
  \end{center}
  \caption{Distribution of the number of reconstructed positrons using the hits in the first two silicon detectors.}
  \label{fig:npositrons}
\end{figure}

The incoming positron position and direction were obtained from the recorded hits in
the two silicon detectors upstream of the target. A vertex constrained fit was performed
using the positron and muons kinematic quantities. This fit first of all allows improving the resolution on the track parameters that will be used later on to measure the emittance. Furthermore this procedure allowed a proper handling of events in which more than one positron was reconstructed. In such cases the positron paired to the muon tracks is the one giving the best goodness of fit. 
For illustration purposes, the observed positron multiplicity, as obtained from a representative data sample with looser cuts, is shown in Fig. \ref{fig:npositrons}.

\section{Results}

\subsection{Data to MC comparisons}

In order to test the reconstruction algorithms presented in the previous section several kinematic quantities, as obtained from the final August 2018 data sample, are compared with the Geant4 expectations obtained processing {\tt Babayaga} MC $\mu^+ \mu^-$ events.
The final data sample is formed by events with two well reconstructed muon tracks, each with a measured momentum, and an incoming positron. The position and direction of the positron were measured 
by the silicon planes upstream of the target. Moreover these three tracks should match to a common vertex, the positron annihilation point in the target.
A sample of 61 events fulfilling the above conditions was obtained.

\begin{figure}[htb]
  \begin{center}
    \includegraphics[width=0.45\linewidth]{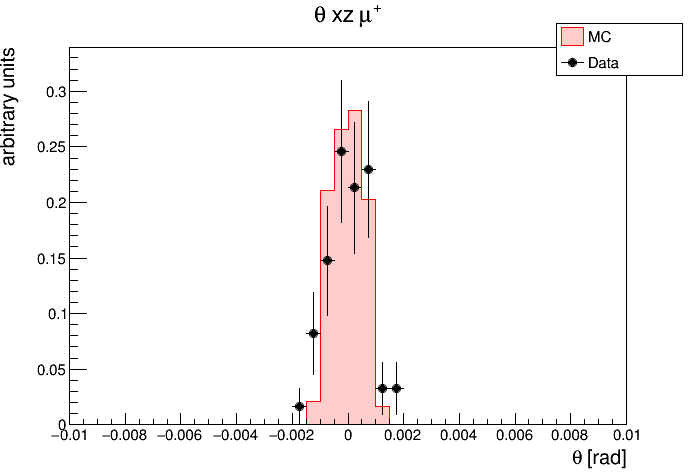}
    \includegraphics[width=0.45\linewidth]{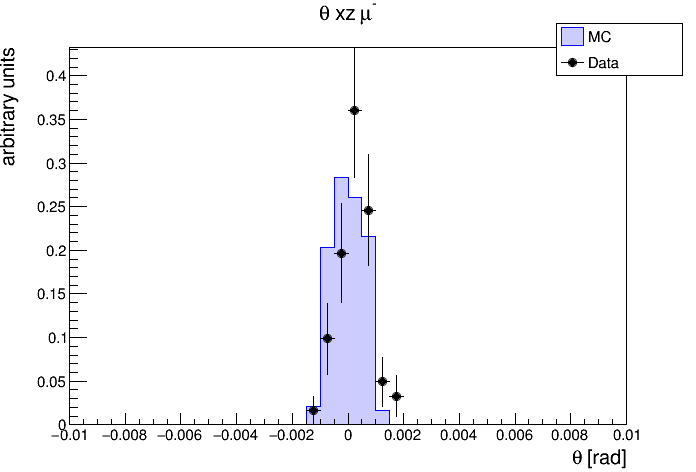}
  \end{center}
  \caption{Reconstructed track angle in the bending plane at the vertex for positively (left) and negatively (right) charged tracks, in events where both are reaching the muon detectors. 
  Data are shown by the dots, simulations by the filled histograms.
  Both data and simulations are normalized to unit area.}
  \label{fig:theta_xz}
\end{figure}

The first kinematic quantity considered is the reconstructed track angle in the bending plane at the vertex.
The corresponding data and simulation shapes are compared in Fig. \ref{fig:theta_xz} separately for positive and negative reconstructed muon tracks. The reconstructed distributions show well collimated tracks, as expected from the Geant4 simulation.

\begin{figure}[htb]
  \begin{center}
    \includegraphics[width=0.45\linewidth]{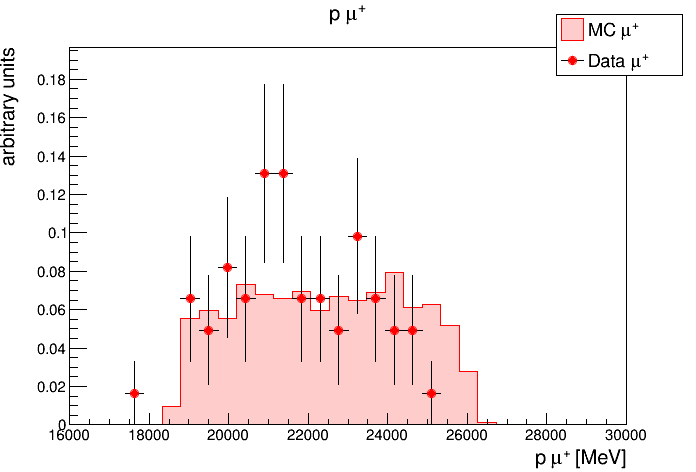}
    \includegraphics[width=0.45\linewidth]{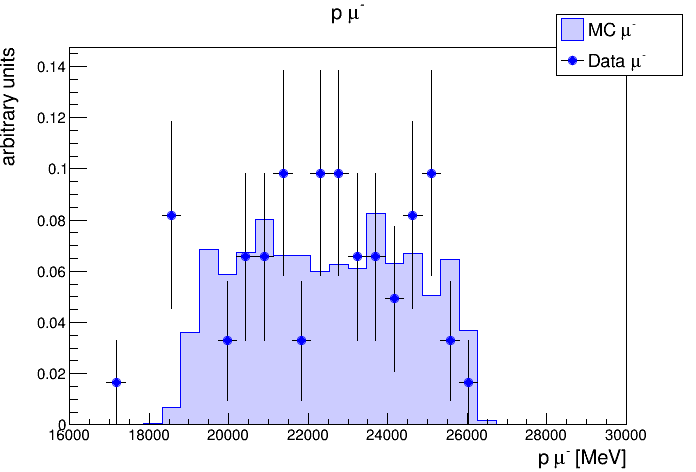}
  \end{center}
  \caption{Reconstructed momentum for positively (left) and negatively (right) charged tracks, in events where both are reaching the muon detectors.
  Data are shown by the dots, simulations by the filled histograms.
  Both data and simulations are normalized to unit area.}
  \label{fig:momenta}
\end{figure}

A similar comparison is performed for the reconstructed tracks momenta in Fig. \ref{fig:momenta}.
As expected from the simulation, the reconstructed distributions are almost flat in the range between 18 and 26 GeV.

\begin{figure}[htb]
  \begin{center}
    \includegraphics[width=0.45\linewidth]{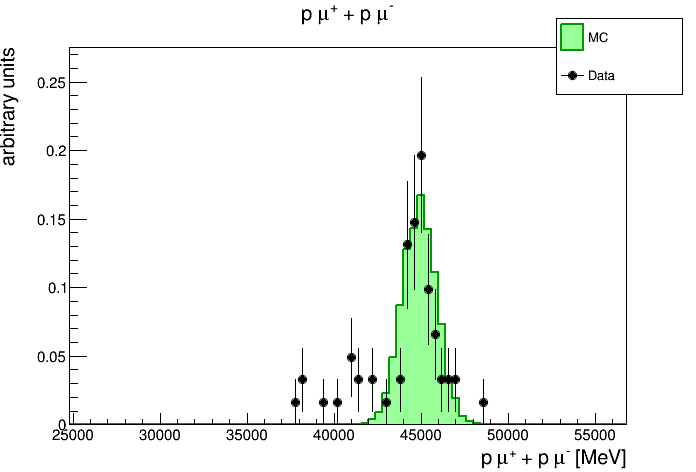}
  \end{center}
  \caption{Reconstructed sum of the muon track momenta. Data are shown by the dots. The filled histogram corresponds to the result of a simulation assuming a $3 \%$ energy resolution on the track momenta. Both data and simulation are normalized to unit area.}
  \label{fig:pTot}
\end{figure}

As a trivial consequence of energy conservation in the process $e^+ e^- \rightarrow \mu^+ \mu^-$
on a target at rest, the sum of the muons momenta should peak at the fixed energy of the incoming
positron beam. The observed experimental spread depends mostly
on the muon track momentum
resolution convoluted with the energy spread of the incoming positron beam. 
The measured shape is shown in Fig. \ref{fig:pTot} and compared to a simulation obtained by smearing the
generated muon track momenta by $3 \%$. This resolution is consistent with what obtained from single muon calibration runs with fixed momentum in the range 18 and 26 GeV. A few data points occur
at values about $10 \%$ smaller than the main peak, around 40 GeV. A similar behaviour is also seen when analyzing single muon calibration runs. Hence events around 40 GeV are more likely to be due to non Gaussian tails in the momentum reconstruction rather than being an unwanted background contribution or arising from a tail in the momentum distribution of the incoming positron beam.

\begin{figure}[htb]
  \begin{center}
    \includegraphics[width=0.45\linewidth]{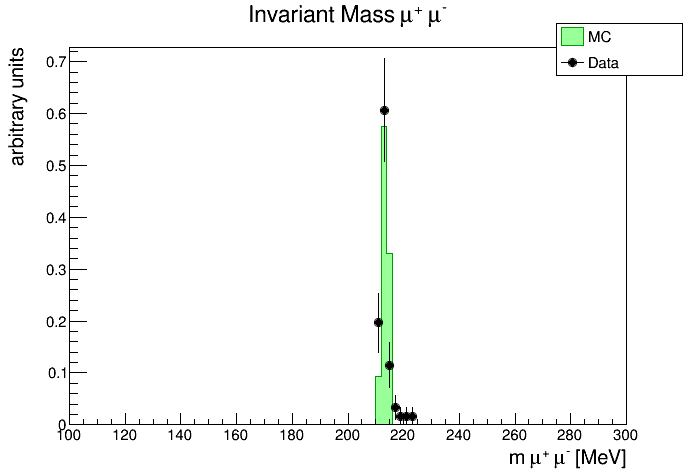}
  \end{center}
  \caption{Invariant mass of the muon track pairs. Data are shown by the dots.
  The filled histogram corresponds to the result of the simulation. Both data and simulation are normalized to unit area.}
  \label{fig:InvMass}
\end{figure}

The last kinematic quantity considered is the invariant mass of the two muon tracks, Fig. \ref{fig:InvMass}. As the reaction $e^+ e^- \rightarrow \mu^+ \mu^-$ is measured at the threshold energy the naive expectation is a
peak at twice the muon mass, i.e. about 212 MeV. This naive expectation, well visible in the selected events, is also confirmed by the simulation results.

Two main sources of inefficiency contribute to the modest
final number of collected events in the August 2018 data taking
period, namely:
\begin{itemize}
    \item the rate limitation of the silicon detectors and calorimeters readout system available, at about 500 Hz, was introducing a large dead time given the observed trigger rate;
    \item the lack of redundancy in the measurements performed in the
    region between the target and the magnet. In order to have well measured tracks, hits in both of the two available detectors had to be requested for both muon tracks. This stringent condition leads to an additional reduction of the efficiency.
\end{itemize}

In September 2018, to reduce the too large dead time of the August runs, an additional scintillator was added, to trigger on events with a track in each arm crossing the corresponding muon chamber, shown in Fig. \ref{fig:tb_setup} with a different symbol.
This action partially mitigated the dead time issue, but due to an hardware misconfiguration the trigger efficiency was very low. Using the trigger-less data recorded by the DT system, the efficiency could be estimated to be as low as $\sim 2 \%$. Once the hardware issue was fixed the estimated efficiency went up to $\sim 10 \%$. A similar estimate of the efficiency was obtained for the August 2018 runs.

A set of data corresponding to a 45 GeV positron beam impinging on a 2 cm thick carbon target recorded after the hardware misconfiguration fix was analysed. Applying all the analysis cuts except the positron--muon pair vertex matching a sample of 157 events was obtained. These reduced to 9 when the positron--muon pair matching condition was applied.
Additional investigations showed that the "geometrical overlap" of the
silicon detectors used in coincidence to tag the incoming positrons was significantly worse in September 2018 w.r.t August. This resulted in a severely limited efficiency for matching the recorded muon pair with the positron that originated it.

\subsection{Raw emittance}

\begin{figure}[htb]
  \begin{center}
    \includegraphics[width=0.45\linewidth]{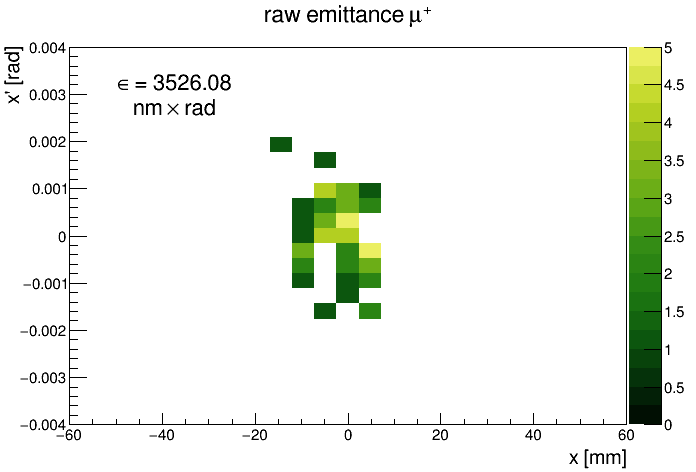}
    \includegraphics[width=0.45\linewidth]{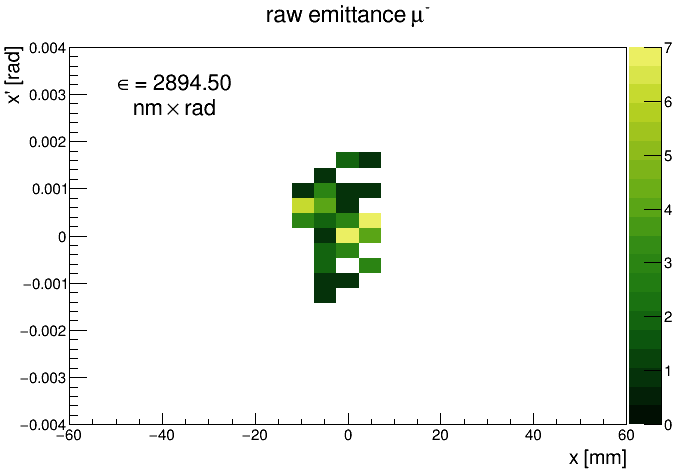}
  \end{center}
  \caption{Raw emittance of the positive (left) and negative (right) tracks. The numerical result is shown as an insert in the plot.}
  \label{fig:raw_emittance}
\end{figure}

The raw emittance is defined in the $x(\mu)$ $x^{\prime}(\mu)$ plane where $x(\mu)$ is the extrapolated position along the $x$-axis of the track at a reference plane taken to be perpendicular to the $z$-axis and 
at a $z$ position corresponding to the target end point. $x^{\prime}(\mu)$ is the corresponding extrapolated local track slope. The observed distributions in the $x(\mu)$ $x^{\prime}(\mu)$ plane, as obtained from the August 2018 data sample, are shown separately for positive and negative muons in Fig. \ref{fig:raw_emittance}. No efficiency corrections are applied.
The numerical values obtained applying Eq. \ref{eq:emittance} are:
\begin{center}
$\epsilon(\mu^+)$ = (3.53 $\pm$ 0.38 (stat.)) $\cdot$ 10$^3$ nm $\times$ rad \\
$\epsilon(\mu^-)$ = (2.89 $\pm$ 0.29 (stat.)) $\cdot$ 10$^3$ nm $\times$ rad \\
\end{center}
where the statistical uncertainty reported above has been obtained from the bootstrap method \cite{bootstrap1,bootstrap2} applied to the 61 events data sample.
The following main sources of systematic uncertainties have been considered: 
\begin{itemize}
    \item variations of the spatial resolution of the tracking detectors: this effect has been investigated repeating several times the analysis of simulated events increasing the resolutions implemented in the simulation by up to 25 \%. This had an impact at the percent level on the raw emittance quoted above.
    \item uncertainty on $e^+ e^- \rightarrow e^+ e^-$ background contamination: this effect has been investigated looking for muon tracks in positron calibration runs (without target). As a result the $e^+ e^-$ background contamination of the final event sample was estimated to be well below the 1 event level.
\end{itemize}
Hence systematic uncertainties are much smaller with respect to the statistical error and neglected.

Corresponding predictions have been obtained from the MC simulation.
Events were generated using an incoming positron beam with the same kinematic properties, in terms of spatial distribution and divergence, as measured in the data with the two silicon detectors upstream of the target. These have been reported in Sec. \ref{sec:analytical}.
Varying the boundaries of the flat spatial distribution or the divergence leads to uncertainties in the predicted raw emittance of about 5 \%.
Averaging the $\mu^+$ and $\mu^-$ simulation results leads to a predicted value of:
\begin{center}
(2.76 $\pm$ 0.15 (modeling)) $\cdot$ 10$^3$ nm $\times$ rad,
\end{center}
in fair agreement with the experimentally measured values.

\subsection{$x$ and $x^{\prime}$ distributions}

The measured raw emittance values are so
large because a very broad distribution in $x$ and $y$ was chosen for the incoming positron beam.
This choice was driven in order to minimize the chance of having two nearby positrons and assigning the wrong one to the measured $\mu^+ \mu^-$ pair.
This resulted in an almost uniform positron distribution in the $x$ range between $-$10 mm and $+$10 mm.
This almost uniform $x(\mu)$ distribution is still visible in Fig. \ref{fig:raw_emittance}, at the reference plane $z$ position. 
Hence the $x(\mu)$ and $x^{\prime}(\mu)$ values that corresponds to the raw emittance have to be corrected by:
\begin{center}
$x = x(\mu) - x(e^+)$ \\
$x^{\prime} = x^{\prime}(\mu) - x^{\prime}(e^+)$
\end{center}
where $x(e^+)$ and $x^{\prime}(e^+)$ are the positions and local slopes obtained from the geometrical extrapolation of the incoming positron to the reference plane.

Simulation studies using generated Geant4 quantities show a width of the $x$ distribution around
30 $\mu$m. At the reconstruction level the observed width that can be achieved with the available experimental setup is significantly larger, around 100 $\mu$m. Hence no useful measurement of the $x$ distribution can be performed.

\begin{figure}[htb]
  \begin{center}
    \includegraphics[width=0.45\linewidth]{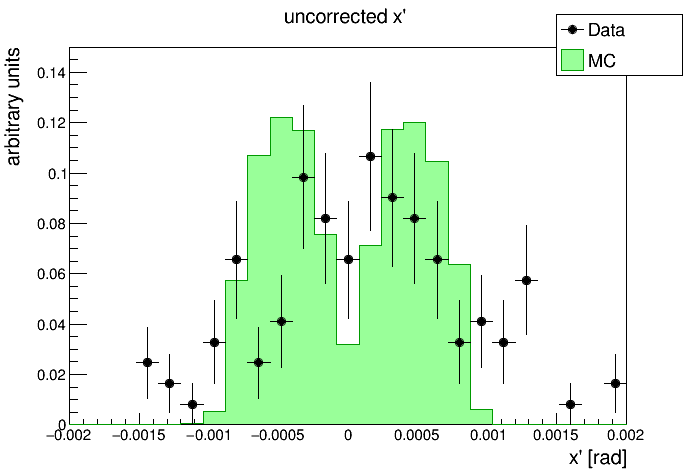}
  \end{center}
  \caption{$x^{\prime}$ distribution: data are shown by the dots,
  the continuous histogram corresponds to the result of the simulation. Both data and simulation are normalized to unit area.}
  \label{fig:xprime}
\end{figure}

A better accuracy is reached on the $x^{\prime}(\mu)$ and $x^{\prime}(e^+)$ measurements.
The $x^{\prime}$ distribution measured in data is shown in Fig. \ref{fig:xprime} together 
with the corresponding simulation result. A fair agreement is found.
The drop around $x^{\prime} = 0$, well visible both in data and simulations, is due to the fact that a fraction of the
events have small values of $x^{\prime}$ for both muon tracks. A small $x^{\prime}$ value corresponds to a small slope. Hence these tracks, produced at the same vertex, will travel almost parallel up to the magnetic field region. 
Due to this track overlap the track reconstruction efficiency is lower for this particular class of events.
The determination of a reliable efficiency correction to account 
for this effect is not possible given the very limited data sample available to validate it.

The contribution to the angular spread $x'$ due to the material present in the path of the muons between the target exit point and the last silicon detector before the magnet has been studied using simulated muon tracks.
The RMS value of the $x'$ distribution has been found to be
0.04 mrad and hence negligible with respect to the width of the
measured $x'$ distribution\footnote{The vacuum pipe reduces the
angular spread by a factor $\sim$ 2.}.

\section{Conclusions}

The $e^+ e^- \rightarrow \mu^+ \mu^-$ annihilation process at threshold energy has been studied with particular emphasis on the kinematic properties of the final state muons. Several unexpected difficulties during the acquisition campaigns lead to a yield of 61 events. 
However for these events both the incoming positron and the outgoing muons are well measured and correlated. So these can be used to get a first estimate of the raw emittance that, within the large data statistical uncertainty, is consistent with analytical calculations and more detailed simulations of the full detector setup. These data also allowed a comparison of the observed uncorrected $x^{\prime}$ distribution to simulations. A fair agreement is found.
In order to subtract the large emittance of the incoming positron beam from the measured raw muon emittance i.e. to provide a measurement of the achievable muon beam emittance
independent from the incoming positron beam characteristics, more accurate tracking devices, alignment infrastructures and more efficient trigger and readout systems will be needed. 
Such a setup could then be used for measurements of the $e^+ e^- \rightarrow \mu^+ \mu^-$ production cross section near threshold. Several accelerator technologies studies, like the development of a muon collider with the LEMMA injection scheme, could benefit from these measurements.

\section*{Acknowledgements}

We would like to warmly thanks the SPS staff and the Large Scale Metrology group, in particular Henrik Wilkens and Nikolaos Charitonidis, for their support during installation and data taking.
This work has been partially supported by the ERC CoG GA 615089 CRYSBEAM.

\end{document}